\documentclass[twoside,12pt]{article}
\usepackage{epsfig}

\def\Journal#1#2#3#4{{#1} {#2} (#4) #3 }

\def\NPA{{\em Nucl. Phys.} A}

\def\PRO{{\em Prog. Theor. Phys.}}

\def\PLB{{\em Phys. Lett.} B}

\def\PRL{\em Phys. Rev. Lett.}

\def\PREP{\em Phys. Rep.}

\def\PRD{{\em Phys. Rev.} D}
\def\PRC{{\em Phys. Rev.} C}

\def\ZPA{{\em Z. Phys.} A}

\newcommand{\be}{\begin{equation}}
\newcommand{\ee}{\end{equation}}
\newcommand{\bea}{\begin{eqnarray}}
\newcommand{\eea}{\end{eqnarray}}

\begin{document}

\title{ \vspace{1cm} Structure of exotic nuclei around double shell closures}
\author{A.\ Covello, L. \ Coraggio, A. \ Gargano, N. \ Itaco  
\\
Dipartimento di Scienze Fisiche, Universit\`a di Napoli Federico II, \\and Istituto Nazionale di Fisica Nucleare, \\
Complesso Universitario di Monte S. Angelo, Via Cintia,  I-80126 Napoli, Italy
\\ }

\maketitle
\begin{abstract} 

In this paper, we first give a brief review of the theoretical framework for microscopic shell-model calculations starting from the free nucleon-nucleon potential. In this context, we discuss the use of the low-momentum nucleon-nucleon interaction $V_{\rm low-k}$ in the derivation of the shell-model effective interaction and emphasize its practical value as an alternative to the Brueckner $G$-matrix method. Then, we present some results of our current study of exotic nuclei around doubly magic $^{132}$Sn, which have been obtained starting from the CD-Bonn potential renormalized by use of the $V_{\rm low-k}$ approach. Attention is focused on the nuclei $^{134}$Te, $^{134}$Sn, and $^{136}$Te, in particular on the latter which is a direct source of information on the proton-neutron effective interaction in the $^{132}$Sn region. Comparison shows that our results are in very good agreement with the available experimental data.
 
\end{abstract}
\section{Introduction}

The study of exotic nuclei in the regions of shell closures is currently drawing much attention. In this context, of special interest are nuclei in the close vicinity to doubly magic $^{100}$Sn and $^{132}$Sn. The experimental study of these nuclei is very difficult, but in recent years new facilities and techniques have made some of them accessible to spectroscopic studies. Today, the development of radioactive ion beams opens up the prospect of gaining substantial information on   $^{100}$Sn and $^{132}$Sn neighbors, which should allow to explore the evolution of the shell structure when approaching the proton and neutron drip lines. This makes it very interesting and timely to perform shell-model calculations to try to explain the available data as well as to make predictions which may stimulate, and be helpful to, future experiments. 

On these grounds, we have recently studied \cite{Coraggio02a,Coraggio04,Coraggio05,Coraggio06}  several nuclei around $^{100}$Sn and $^{132}$Sn within the framework of the shell model employing realistic effective interactions derived from the CD-Bonn nucleon-nucleon ($NN$) potential \cite{Machleidt01}. 
A main difficulty encountered in this kind of calculations is the strong short-range repulsion contained in the bare $NN$ potential $V_{NN}$, which prevents its direct use in the derivation of the shell-model effective interaction $V_{\rm eff}$.
As is well known, the traditional way  to overcome this difficulty is the Brueckner
$G$-matrix method. Instead, in the calculations mentioned above we have made use of a new approach  \cite{Bogner02} which consists in deriving from $V_{NN}$ a low-momentum potential, $V_{\rm low-k}$, that preserves the deuteron binding energy and scattering phase shifts of $V_{NN}$ up to a certain cutoff momentum $\Lambda$. This is a smooth potential which can be used directly to derive $V_{\rm eff}$, and it has been shown \cite{Bogner02,Covello02,Covello03} that it provides an advantageous alternative to the use of the $G$ matrix. 

The main aim of this paper is to give a brief survey of the $V_{\rm low-k}$ approach
and show some results of its practical application in realistic shell-model calculations. In particular, we shall focus attention on the three even-mass nuclei  $^{134}$Te, $^{134}$Sn, and $^{136}$Te. The study of these nuclei, which have as valence particles  two protons, two neutrons, and two protons plus two neutrons, respectively, provides a stringent test of our realistic effective interaction in the $^{132}$Sn region.

The paper is organized as follows. In Sec. 2 we give an outline of the derivation
of $V_{\rm low-k}$ and $V_{\rm eff}$. In Sec. 3 we first give some details of the calculations, and then present and discuss our results comparing them with the  available experimental data. Section 4 contains some concluding remarks.

\section{Outline of Theoretical Framework}

The shell-model effective interaction $V_{\rm eff}$ is defined, as usual, in the following way. In principle, one should solve a nuclear many-body Schr\"odinger equation of the form 
\be
H\Psi_i=E_i\Psi_i ,
\ee
with $H=T+V_{NN}$, where $T$ denotes the kinetic energy. This full-space many-body problem is reduced to a smaller model-space problem of the form
\vspace{-.1cm}
\be
PH_{\rm eff}P \Psi_i= P(H_{0}+V_{\rm eff})P \Psi_i=E_iP \Psi_i .
\ee
\noindent Here $H_0=T+U$ is the unperturbed Hamiltonian, $U$ being an auxiliary potential introduced to define a convenient single-particle basis, and $P$ denotes the projection operator onto the chosen model space.

As pointed out in the Introduction, we ``smooth out" the strong repulsive core contained in the bare $NN$ potential $V_{NN}$ by constructing a low-momentum  potential
$V_{\rm low-k}$. This is achieved by integrating out the high-momentum modes of $V_{NN}$ down to a cutoff momentum  $\Lambda$. This integration is carried out with the requirement that the deuteron binding energy and phase shifts of $V_{NN}$ up to $\Lambda$ are preserved by $V_{\rm low-k}$. This requirement may be satisfied by the following $T$-matrix equivalence approach. We start from the half-on-shell $T$ matrix for $V_{NN}$ 
\begin{equation}
T(k',k,k^2) = V_{NN}(k',k) + \wp \int _0 ^{\infty} q^2 dq  V_{NN}(k',q)
\frac{1}{k^2-q^2} T(q,k,k^2 )\,,
\end{equation}
where $\wp$ denotes the principal value and  $k,~k'$, and $q$ stand for the relative momenta. 
The effective low-momentum $T$ matrix is then defined by

\be
T_{\rm low-k }(p',p,p^2) = V_{\rm low-k }(p',p)   \\
+ \,\wp \int _0 ^{\Lambda} q^2 dq  
V_{\rm low-k }(p',q) \frac{1}{p^2-q^2} T_{\rm low-k} (q,p,p^2)\ ,  \label{eq:ham}
\ee
where the intermediate state momentum $q$ is integrated from 0 to the momentum cutoff 
$\Lambda$ and \linebreak $(p',p) \leq \Lambda$. 
These two $T$ matrices are required to satisfy the condition 
\be
T(p',p,p^2)= T_{\rm low-k }(p',p,p^2) \, ; ~~ (p',p) \leq \Lambda \,.
\ee

The above equations define the effective low-momentum interaction $V_{\rm low-k}$, and it has been shown \cite{Bogner02} that they are satisfied by the solution
\begin{equation}
V_{\rm low-k} = \hat{Q} - \hat{Q'} \int \hat{Q} + \hat{Q'} \int \hat{Q} \int
\hat{Q} - \hat{Q'} \int \hat{Q} \int \hat{Q} \int \hat{Q} + ~...~~,
\end{equation}
which is the well known Kuo-Lee-Ratcliff (KLR) folded-diagram expansion \cite{KLR71,Kuo90}, originally designed for constructing  shell-model effective interactions.
In the above equation $\hat{Q}$ is an irreducible vertex function whose intermediate states are all beyond $\Lambda$ and $\hat{Q'}$ is obtained by removing from $\hat{Q}$ its terms first order in the interaction $V_{NN}$. In addition to the preservation of the half-on-shell $T$ matrix, which implies preservation of the phase shifts, this $V_{\rm low-k}$ preserves the deuteron binding energy, since eigenvalues are preserved by the KLR effective interaction. 
For any value of $\Lambda$, the low-momentum effective interaction of Eq. (6) can be calculated very accurately using iteration methods. Our calculation of $V_{\rm low-k}$ is performed by employing the iterative implementation of the Lee-Suzuki method \cite{Suzuki80} proposed in \cite{Andreozzi96}.

The $V_{\rm low-k}$ given by the $T$-matrix equivalence approach mentioned above is not Hermitian. Therefore, an additional transformation is needed to make it Hermitian. To this end, we resort to the Hermitization procedure suggested in \cite{Andreozzi96}, which makes use of the Cholesky decomposition of symmetric positive definite matrices.  

\begin{figure}[h]
\epsfysize=9.0cm
\begin{center}
\epsfig{file=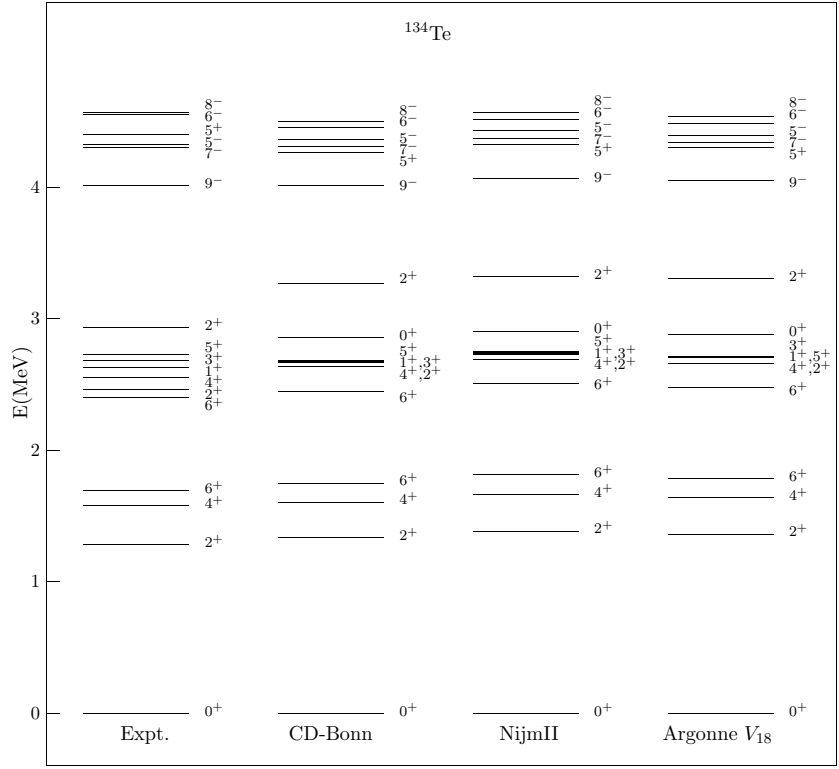,scale=0.7}
\caption{Spectrum of $^{134}$Te. Predictions by various $NN$ potentials are compared with experiment.\label{fig1}}
\end{center}
\end{figure}

Once the $V_{\rm low-k}$ is obtained, we use it, plus the Coulomb force for protons, as input interaction  for the calculation of the matrix elements of $V_{\rm eff}$.
The latter is derived by employing a folded-diagram method, which was previously applied to many nuclei \cite{Covello01} using $G$-matrix interactions. Since $V_{\rm low-k}$ is already a smooth potential, it is no longer necessary to calculate the $G$ matrix. We therefore perform shell-model calculations following the same procedure as described, for instance, in \cite{Jiang92,Covello97}, except that the $G$ matrix used there is replaced by $V_{\rm low-k}$. More precisely, we first calculate the so-called 
$\hat{Q}$-box \cite{Kuo80} including diagrams up to second order in the two-body interaction. The shell-model effective interaction is then obtained by summing up the $\hat{Q}$-box folded-diagram series using the Lee-Suzuki iteration method \cite{Suzuki80}.

Clearly, the starting point of any realistic shell-model calculation is the free $NN$ potential. There are, however, several high-quality potentials, such as Nijmegen I and Nijmegen II \cite{Stoks94}, Argonne $V_{18}$ \cite{Wiringa95}, and CD-Bonn \cite{Machleidt01}, which fit equally well ($\chi^2$/datum $\approx 1$) the $NN$ scattering data up to the inelastic threshold. This means that their on-shell properties are essentially identical, namely they are phase-shift equivalent.

In our shell-model calculations we have derived the effective interaction from the CD-Bonn  potential. This may raise the question of how much our results may depend on this choice of the $NN$ potential. We have verified that shell-model effective interactions derived from phase-shift equivalent $NN$ potentials through the 
$V_{\rm low-k}$ approach do not lead to significantly different results. Here, by way of illustration, we present the results obtained for the nucleus $^{134}$Te. This nucleus has only two valence protons and thus offers the opportunity to test directly the matrix elements of the various effective interactions. In Fig. 1 we show, together with the experimental spectrum, the spectra obtained by using the CD-Bonn, NijmII, and Argonne $V_{18}$ potentials, all renormalized through the $V_{\rm low-k}$ procedure with a cutoff momentum $\Lambda$=2.2 fm$^{-1}$. This value of $\Lambda$ is in accord with the criterion given in \cite{Bogner02}. From Fig. 1 we see 
that the calculated spectra are very similar, the differences between the level energies not exceeding 80 keV. It is also seen that the agreement with experiment is very good for all the three potentials.

\section{Calculations and Results}

We now present and discuss the results of our study of the even-mass nuclei
$^{134}$Te, $^{134}$Sn, and  $^{136}$Te . The calculations have been performed using the OXBASH shell-model code \cite{OXBASH}.

We consider $^{132}$Sn as a closed core 
and let the valence protons occupy the five levels  $0g_{7/2}$, $1d_{5/2}$, 
$1d_{3/2}$, $2s_{1/2}$, and $0h_{11/2}$ of the 50-82 shell, while for the valence neutrons the model space includes the six levels $0h_{9/2}$, $1f_{7/2}$, $1f_{5/2}$, $2p_{3/2}$, $2p_{1/2}$, and  $0i_{13/2}$ of the 82-126 shell.

As mentioned in the previous section, the two-body matrix elements of the effective interaction are derived from the CD-Bonn $NN$ potential renormalized through the 
$V_{\rm low-k}$ procedure with a cutoff momentum $\Lambda=2.2$ fm$^{-1}$.  The computation of the diagrams included in the $\hat{Q}$-box is performed within the harmonic-oscillator basis using intermediate states composed of all possible hole states and particle states restricted  to the five shells above the Fermi surface. The oscillator parameter used is $\hbar \omega = 7.88$ MeV.

As regards the single-proton and single-neutron energies, we have taken them from the experimental spectra of $^{133}$Sb and $^{133}$Sn \cite{NNDC}. In the spectra of these nuclei, however, some single-particle levels are still missing. More precisely, this is the case of the proton $2s_{1/2}$ and neutron $0i_{13/2}$ levels, whose energies have been taken from Refs. \cite{Andreozzi97} and \cite{Coraggio02b}, respectively, where it is discussed how they are determined. 

\begin{figure}[h]
\epsfysize=9.0cm
\begin{center}
\epsfig{file=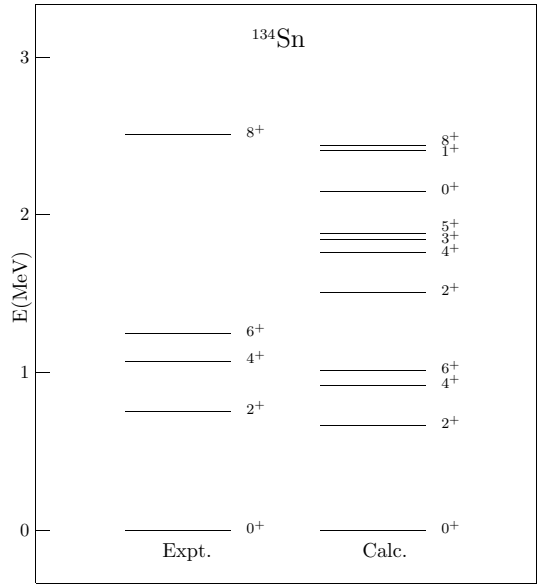,scale=0.7}
\caption {Experimental and calculated spectra of $^{134}$Sn. \label{fig3}}
\end{center}
\end{figure}

In Fig. 2 we compare the experimental \cite{Zhang97,Korgul00} and calculated spectra  of $^{134}$Sn. Wee see that all the observed levels are very well reproduced by the theory, which also predicts, in between the $6^+$ and $8^+$ states, the existence of other five states with $J^{\pi}=0^+$, $2^+$, $3^+$, $4^+$, and $5^+$. These states  all arise from  configurations involving the $p_{3/2}$ single-neutron level and are then  expected to lie below the $8^+$ state which is the highest-spin member of the 
$f_{7/2} h_{9/2}$ multiplet. 
Is  is not surprising that they have not been observed in the experiment of 
Ref. \cite{Korgul00}, where $^{134}$Sn was studied through the prompt-$\gamma$ 
spectroscopy of fission fragments. 

As  regards $^{134}$Te, the spectrum obtained using the CD-Bonn potential has already been compared with experiment in the previous section (see Fig. 1), where we discussed the dependence of shell-model results on the $NN$ potential used as input. In that context, we noted that the agreement between theory and experiment is very good.
From Fig. 1, where all calculated and observed levels are reported 
up to about 3.2 MeV, we see that a one-to-one correspondence exists between the experimental and theoretical spectra, the only exception being a 
$0^+$ state  predicted around 2.5 MeV excitation energy.
Above 3.2 MeV  we have only reported  the observed levels which have received a spin-parity assignment and the calculated ones which are candidates for them. 

\begin{figure}[h]
\epsfysize=9.0cm
\begin{center}
\epsfig{file=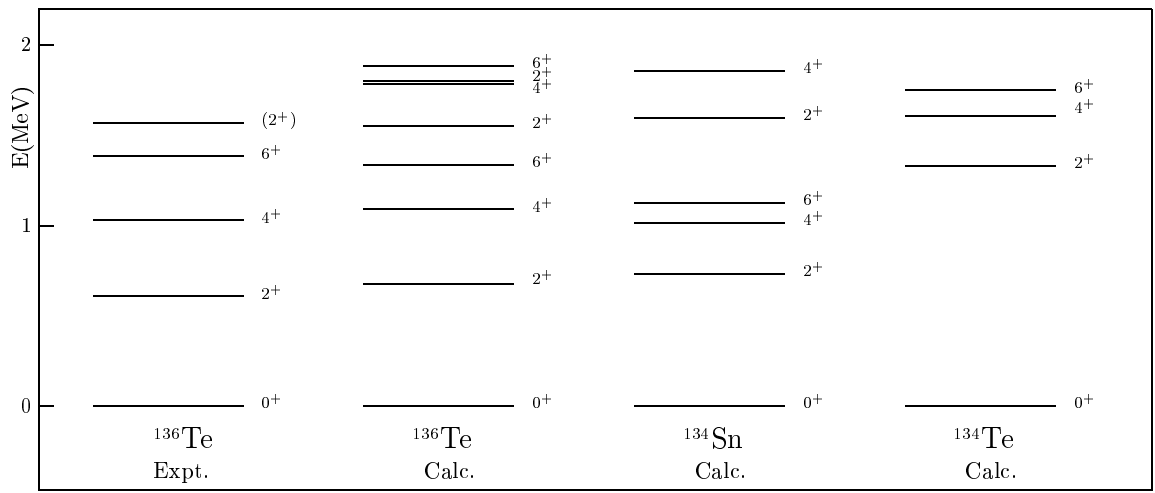,scale=0.8}
\caption {Experimental and calculated spectra of $^{136}$Te. Also shown are the calculated spectra of $^{134}$Sn and $^{134}$Te (see text for comments). 
\label{fig2}}
\end{center}
\end{figure}

Let us now come to the two-proton, two-neutron nucleus $^{136}$Te, which in recent 
years has been the subject of great experimental and theoretical interest \cite{Radford02,Terasaki02,Shimizu04,Brown05}. Particular attention 
has been focused on the first 
$2^+$ state, which shows a significant drop in energy as compared to $^{134}$Te, and on  the   
$B(E2; 0^{+} \rightarrow 2^{+})$  value, recently measured using Coulomb excitation of radioactive 
ion beams \cite{Radford02}. In fact, these data seem to show evidence of an anomalous behavior in the Te isotopes above the $N=82$ shell closure.

Here, we present our results for $^{136}$Te and discuss the nature of  its states 
showing how they are formed as a result of the neutron-proton effective interaction starting from the two-neutron and two-proton systems.
In Fig. 3 the experimental \cite{NNDC} and calculated spectra of   $^{136}$Te up
to about 2 MeV excitation energy are shown and we see that the
experimental spectrum  is very well reproduced by the theory. In the same figure 
we also report the calculated 
levels of $^{134}$Sn and $^{134}$Te up to the same energy.  When comparing the three calculated spectra,
it is worth noting 
that the spectrum of $^{136}$Te may be seen as resulting from the sum of the $^{134}$Sn and $^{134}$Te spectra.
There is only a missing third $4^+$ state, which, lying at about 2.2 MeV, is not included in 
the figure. In this context, we have found it interesting to relate  the wave functions of
the low-lying states of   $^{136}$Te to those of   $^{134}$Sn and $^{134}$Te. 

Let us consider the ground state and the $2^+$, $4^+$, and $6^+$ yrast states.
Their wave functions can be written as

\begin{equation}
|^{136}{\rm Te; g.s} \rangle = 0.85 |^{134}{\rm Sn; g.s.} \rangle |^{134}{\rm Te; g.s.} \rangle  + \cdots ,\\
\end{equation}

\begin{equation}
|^{136}{\rm Te}; 2_{1}^{+} \rangle = 0.73 |^{134}{\rm Te}; g.s.  \rangle |^{134}{\rm Sn};  2_{1}^{+} \rangle  + 
0.36 |^{134}{\rm Sn}; g.s.  \rangle  |^{134}{\rm Te};  2_{1}^{+} \rangle  + \cdots ,\\
\end{equation}

\begin{equation}
|^{136}{\rm Te};  4_{1}^{+} \rangle = 0.71 |^{134}{\rm Te} ; g.s. \rangle |^{134}{\rm Sn}; 4_{1}^{+}\rangle  + 
0.28 |^{134}{\rm Sn}; g.s. \rangle |^{134}{\rm Te};  |4_{1}^{+} \rangle  + \cdots ,\\
\end{equation}

\begin{equation}
|^{136}{\rm Te}; 6_{1}^{+} \rangle = 0.78 |^{134}{\rm Te}; g.s.  \rangle |^{134}{\rm Sn};  6_{1}^{+}\rangle  - 
0.21 |^{134}{\rm Sn}; g.s.  \rangle |^{134}{\rm Te};  6_{1}^{+} \rangle  + \cdots ,\\
\end{equation}
where ``$\cdots$'' means other minor components. As it might have been  supposed by
an inspection of Fig.~3, we see that all the above states are dominated, 
by two neutrons excitations, the correlations between the two protons being just the same as in the 
$^{134}$Te ground state.  This may be traced to the fact that the proton-neutron interaction is 
not strong enough to overcome the energy difference
between the two corresponding yrast states in  $^{134}$Sn and $^{134}$Te. 

A similar analysis was performed in \cite{Shimizu04} for the first two $2^+$ states and a result 
similar to ours was found for the yrast one. 
It should be mentioned, however, that the study of Ref. \cite{Shimizu04}  
was based on a paring plus quadrupole-quadrupole interaction and predicts for the $2_{2}^{+}$ state
a structure different from the one we find.
Our wave functions for the second and third $2^+$ states are 

\begin{equation}
|^{136}{\rm Te};  2_{2}^{+} \rangle = 0.61 |^{134}{\rm Te}; g.s.  \rangle |^{134}{\rm Sn};  2_{2}^{+}\rangle  - 
0.42 |^{134}{\rm Te}; g.s. \rangle  |^{134}{\rm Sn};  2_{1}^{+} \rangle  + \cdots ,\\
\end{equation}

\begin{equation}
|^{136}{\rm Te};  2_{3}^{+} \rangle = 0.78 |^{134}{\rm Sn}; g.s.  \rangle |^{134}{\rm Te};  2_{1}^{+} \rangle  - 
0.31 |^{134}{\rm Te}; g.s.  \rangle  |^{134}{\rm Sn};   2_{1}^{+} \rangle  + \cdots .\\
\end{equation}
Actually, we find that the $2_{2}^+$ state arises from the $2_{2}^+$  of $^{134}$Sn, while the 
$2_{3}^+$ state, dominated by the two-proton $2_{1}^+$  excitation, is a  so-called ``mixed-symmetry'' state.
The relevance of this kind of states is discussed in detail in  
\cite{Shimizu04}, where the ``mixed-symmetry'' state was, however, identified with the second $2^+$ state.
More experimental data are needed to clarify this point.

\begin{table}[h]
\caption{Experimental and calculated $B(E2)$ values (in W.u.) for $^{136}$Te, $^{134}$Sn, and $^{134}$Te.}
\begin{center}
\begin{tabular} {c|cc|cc|cc}
\hline  
&&&& \\ [-2mm]
{$J^\pi_{i} \rightarrow J^\pi_{f}$}  & \multicolumn{2} {c|} {~~~$^{136}$Te} &\multicolumn{2}{c|}{~~~$^{134}$Sn} &\multicolumn{2}{c} {~~~$^{134}$Te}  \\  
   &
{    Expt. } &
{   Calc.} &
{  Expt.} & 
{    Calc. } &
{   Expt.} &
{    Calc.}  \\
&&&& \\ [-2mm]
\hline 
&&&&& \\ [-2mm]
$0^{+} \rightarrow 2^{+}$ & $25 \pm 4$ & 44 &  $7 \pm 1$       & 8    & $24 \pm 3$     & 20 \\
$4^{+} \rightarrow 2^{+}$ &           & 11 &                   & 1.6  & $4.3 \pm 0.40$  & 4.3 \\
$6^{+} \rightarrow 4^{+}$ &           & 7.5 & $0.89 \pm 0.17$ & 0.81  & $2.05 \pm 0.04$ & 1.9 \\ [2mm]
\hline
\end{tabular}
\end{center}
\end{table}

To conclude this section, we report in Table 1 several  experimental \cite{NNDC,Radford02,Beene04} and 
calculated $E2$ transition rates for $^{134}$Sn, $^{134}$Te, and
$^{136}$Te.  The calculated values have been obtained using effective proton and neutron charges of 1.55$e$ and
0.70$e$, respectively \cite{Andreozzi97,Coraggio02b}.  We see that our results are in very good agreement with the experimental
values, the only significant discrepancy occurring for the $B(E2; 0^{+} \rightarrow 2^{+})$ in $^{136}$Te.
It should be mentioned, however, that this is likely not to be the case, as a new measurement \cite{Baktash06} yields a value which is $50 \%$ 
higher than that reported in Table 1.

\section{Concluding Remarks}

We have presented here the results of a shell-model study of neutron-rich nuclei in the close vicinity to $^{132}$Sn, focusing attention on the three nuclei $^{134}$Te, $^{134}$Sn, and $^{136}$Te, which are most appropriate for a stringent test of the nucleon-nucleon effective interaction in the $^{132}$Sn region. 

Our two-body effective interaction  has been derived by means of a $\hat Q$-box 
folded-diagram method from the CD-Bonn $NN$ potential renormalized by use of the $V_{\rm low-k}$ approach. In this regard, it should be emphasized that the $V_{\rm low-k}$'s extracted from various modern $NN$ potentials give very similar results in shell-model calculations, suggesting the realization of a nearly unique low-momentum $NN$ potential.

We have shown that our results are all in very good agreement with experiment. In particular, our calculations reproduce quite accurately the $B(E2;0^+ \rightarrow 2_1^+)$ values that have been recently measured through Coulomb excitation of radioactive ion beams.

\section*{Acknowledgments}

This work was supported in part by the Italian Ministero dell'Istruzione, dell'Universit\`a e della Ricerca (MIUR).


\begin{thebibliography}{99}
\itemsep -2pt 

\bibitem{Coraggio02a} L. Coraggio, A. Covello, A. Gargano, N. Itaco, and T. T. S. Kuo, \Journal{\PRC} {66} {064311} {2002}.
\bibitem{Coraggio04} L. Coraggio, A. Covello, A. Gargano, and N. Itaco, \Journal
{\PRC}{70} {034310} {2004}.
\bibitem{Coraggio05} L. Coraggio, A. Covello, A. Gargano, and N. Itaco, \Journal{\PRC}
{72} {057302} {2005}.
\bibitem{Coraggio06} L. Coraggio, A. Covello, A. Gargano, and N. Itaco, \Journal{\PRC}
{73} {031302(R)} {2006}.
\bibitem{Machleidt01} R. Machleidt, \Journal{\PRC} {63} {024001} {2001}.
\bibitem{Bogner02} S. Bogner, \ \ T. T. S. Kuo, \ \ L. Coraggio, \ \ A. Covello, \ \ and N. Itaco,  \Journal{\PRC}{65}{051301(R)} {2002}.
\bibitem{Covello02} A. Covello, L. Coraggio, A. Gargano, N. Itaco, and T. T. S. Kuo,\
in {\em Challenges of Nuclear Structure}, ed. A. Covello (World Scientific, Singapore, 2002), p. 139.
\bibitem{Covello03} A. Covello, in {\em  Proc. Int. School of Physics ``E. Fermi", Course CLIII}, eds. A. Molinari, L. Riccati, W. M. Alberico, and M. Morando (IOS Press, Amsterdam, 2003), p. 79.
\bibitem{KLR71} T. T. S. Kuo, S. Y. Lee, and K. F. Ratcliff, \Journal{\NPA} {176} {65}
{1971}.
\bibitem{Kuo90} T. T. S. Kuo and E. Osnes, {\it Lecture Notes in Physics}, Vol. ~ 364 (Springer-Verlag, Berlin, 1990).
\bibitem{Suzuki80} K. Suzuki, S. Y. Lee,  \Journal{\PRO} {64} {2091} {1980}.
\bibitem{Andreozzi96} F. Andreozzi,  \Journal{\PRC} {54} {684} {1996}.
\bibitem{Covello01} A. Covello, L. Coraggio, A. Gargano, and N. Itaco, {\it Acta Phys. Pol.}  B 32 (2001) 871, and references therein.
\bibitem{Jiang92} M. F. Jiang, R. Machleidt, D. B. Stout, and T. T. S. Kuo, 
\Journal{\PRC} {46} {910} {1992}. 
\bibitem{Covello97} A. Covello, F. Andreozzi, L. Coraggio, A. Gargano , T. T. S. Kuo, and A. Porrino, {\it Prog. Part. Nucl. Phys.} 38,  (1997) 165.
\bibitem{Kuo80} T. T. S. Kuo and E. M. Krenciglowa, \Journal{\NPA} {342} {454} {1980}.
\bibitem{Stoks94} V. G. J. Stoks, R. A. M. Klomp, C. P. F. Terheggen, and J. J. de Swart,  \Journal{\PRC} {49} {2950} {1994}.
\bibitem{Wiringa95} R. B.Wiringa, V. G. J. Stoks, and R. Schiavilla, \Journal{\PRC}
{51} {38} {1995}.
\bibitem{OXBASH} B. A. Brown, A. Etchegoyen, and W. D. M. Rae, {\em The computer code OXBASH}, MSU-NSCL, Report No. 524.
\bibitem{NNDC} Data extracted using the NNDC On-line Data Service from the ENSDF database, version of November 3, 2006.
\bibitem{Andreozzi97} F. Andreozzi, L. Coraggio, A. Covello, A. Gargano, T. T. S. Kuo, and  A. Porrino, \Journal {\PRC} {56} {R16} {1997}.
\bibitem{Coraggio02b} L. Coraggio, A. Covello, A. Gargano, and N. Itaco, 
\Journal{\PRC} {65} {051306(R)} {2002}.
\bibitem{Zhang97} C. T. Zhang {\it et al.}, \Journal {\ZPA} {358} {9} {1997}.
\bibitem{Korgul00} A. Korgul {\it et al.}, {\em Eur. Phys. J.} A 7 (2000) 167.
\bibitem{Radford02} D. C. Radford {\it et al.}, \Journal {\PRL} {88} {222501} {2002}.
\bibitem{Terasaki02} J. Terasaki, J. Engel, W. Nazarewicz, and M. Stoitsov,
\Journal {\PRC} {66} {054313} {2002}.
\bibitem{Shimizu04} N. Shimizu, T. Otsuka, T. Mizusaki, and M. Honma, \Journal {\PRC}
{70} {054313} {2004}. 
\bibitem{Brown05}B. A. Brown, N. J. Stone, R. J. Stone, I. S. Towner, and M. Hjorth-Jensen, \Journal {\PRC}  {71} {044317}  {2005}.
\bibitem{Baktash06} C. Baktash (private communication).
\bibitem{Beene04} J. R. Beene {\it et al.} \Journal {\NPA} {746} {471c} {2004}.


\end{thebibliography}
\end{document}